# Formula for red-shift of light signals coming from distant galaxies

Jian-Miin Liu

Department of Physics, Nanjing University

Nanjing, The People's Republic of China

On leave, E-mail address: liu@phys.uri.edu

## Abstract

Relying on the obtained results in Ref.[1], we derive the formula relating the red-shift of light signals coming from distant galaxies to the distance of these galaxies from us and the time of detecting of these light signals. The red-shift coefficient, instead of the Hubble parameter, is introduced. It varies with time and positive at all times. Its nowadays value equals the Hubble parameter. It increases forever as time is running from the past to the future. The derived formula enables us to estimate the nowadays increasing rate of the red-shift coefficient, which is nothing but the nowadays value of the "acceleration of the expansion of the Universe".

## 1. Introduction

In the Earth-related coordinate system { $x^0, r, \theta, \phi$ }, $x^0 = ct$, we reconstructed the standard model of cosmology based on the assumption of the cosmological principle and the perfect gas (or fluid) [1]. We exactly solved Einstein's field equation involved,

$$R^\mu_\nu - \frac{1}{2} R^\lambda_\lambda \delta^\mu_\nu = 8\pi G T^\mu_\nu + \Lambda \delta^\mu_\nu, \quad \mu = \nu = 0,1,2,3, \tag{1}$$

with the energy-momentum tensor

$$T^\mu_\nu = \begin{cases} \rho, & for \mu = \nu = 0, \\ -p\delta^\mu_\nu, & for \mu = \nu = 1,2,3, \\ 0, & otherwise, \end{cases} \tag{2}$$

where $c$ is the speed of light in vacuum, G is the Newtonian gravitational constant, $\Lambda$ is the cosmological constant, $R^\mu_\nu$ is the Ricci curvature tensor, $R^\lambda_\lambda$ is the space-time curvature, and $\rho$ and $p$ are respectively the mass-density and pressure of the matter of the Universe which presents as a perfect gas. The exact solution consists of three parts:

$$\Lambda = 0, \tag{3a}$$

$$\rho + p = 0, \tag{3b}$$

$$ds^2 = c^2 dt^2 - \exp[2\sqrt{\frac{8\pi\rho G}{3}}t]\{dr^2 + r^2 d\theta^2 + r^2 \sin^2\theta d\phi^2\}, \tag{3c}$$

where $t$ is relative to the SC-moment at which the geometric structure of space-time of the Universe is of the Minkowskian [1]. Eq.(3a) specifies the cosmological constant to be zero. Eq.(3b) exhibits the equation of state for the matter of the Universe. Eq.(3c) is the line element for space-time of the Universe.

The line element of space-time of the Universe varies with time, actually with only time. We can represent the line element in the Earth-related Cartesian coordinate system { $x^0, x^r$ }, $r = 1,2,3$,

$$x^1 = r\sin\theta\cos\phi, \quad x^2 = r\sin\theta\sin\phi, \quad x^3 = r\cos\theta,$$

and clearly see this as the represented line element is

$$ds^2 = c^2 dt^2 - \exp[2\sqrt{\frac{8\pi\rho G}{3}}t]\delta_{rs}dx^r dx^s. \tag{4}$$

The time-evolution of the line element surely affects some physical processes that occur in large region of space. In this paper, we are concerned with the red-shift of light signals coming from distant galaxies [2-8].

## 2. The Earth-related coordinate system

Einstein made an exquisite definition, from the measurement point of view, for inertial coordinate system: "in a given inertial frame of reference the coordinates mean the results of certain measurements with rigid (motionless) rods, a clock at rest relative to the inertial frame of reference defines a local time, and the local time at all points of space, indicated by synchronized clocks and taken together, give the time of this inertial frame of reference."[9]. As defined, an observer binding to an inertial frame of reference can employ his motionless-rigid rods and motionless-synchronized clocks in the so-called "motionless-rigid rod and motionless-synchronized clock" measurement method to measure space and time intervals. By using this "motionless-rigid rod and motionless-synchronized clock" measurement method, the observer sets up his inertial coordinate system [10-12]. This definition is of course suitable for the case of no gravitational field. In the presence of a gravitational field, the inertial frame of reference is no longer inertial and a different set of motionless-rigid rods and motionless-synchronized clocks will be called for and employed by the observer. It is therefore our understanding that an observer binding to the Earth-related inertial (at least approximately inertial) frame of reference in which the Earth is always at rest still can, in the presence of the gravitational field, set up a coordinate system by using the motionless-rigid rods and motionless-synchronized clocks existing in the gravitational field. We call this coordinate system the Earth-related coordinate system. It is not inertial unless the gravitational field is turned off. In our case of the gravitational field of the Universe, we can set up the Earth-related coordinate system by use of the motionless-rigid rods and motionless-synchronized clocks existing in the gravitational field of the Universe.

Since it is the motionless-rigid rods and motionless-synchronized clocks existing in the gravitational field of the Universe that we employ in our cosmological measurements, the measurement data are all available in the Earth-related coordinate system. On the other hand, we did [1] and we now do theoretical calculations also in the Earth-related coordinate system. The comparison between calculation results and measurement data can be therefore made directly in the Earth-related coordinate system.

To get the equation of motion for light propagation, we can set $ds^2 = 0$ in the line element Eq.(3c). Doing so, we have

$$c^2 dt^2 = \exp[2\sqrt{\frac{8\pi\rho G}{3}}t]\{dr^2 + r^2 d\theta^2 + r^2\sin^2\theta d\phi^2\}.$$

For purely radial propagation of light, it reduces to

$$\exp[-2\sqrt{\frac{8\pi\rho G}{3}}t]c^2 dt^2 = dr^2. \tag{5}$$

This equation is a base for our discussions below.

### 3. **Red-shift of light signals from distant galaxies**

Suppose a light signal of frequency $\nu_1$ is emitted at time $t_1$ on a galaxy located at ($r, \theta, \phi$). This signal is later detected by us on the Earth at time $t$ and determined to have frequency $\nu$. It is then compared to a light signal of the same type, produced or emitted at time $t$ on the Earth, whose frequency is $\nu_0$. Experimental measurements reveal a red-shift,

$$\nu < \nu_0 \text{ or } z \equiv \frac{\nu_0 - \nu}{\nu} > 0 \,. \tag{6}$$

According to Hubble, the red-shift is proportional to distance $r$ of the galaxy from us,

$$cz = Hr \,, \tag{7a}$$

where $H$ is the Hubble parameter or constant. Still according to Hubble, the red-shift is due to the Doppler effect, and hence

$$y \approx Hr \,, \tag{7b}$$

where $y$ is the recessional speed of the galaxy from us. Both Eqs.(7a-b) are called the Hubble law [2-8, 13-18].

The emissions of two light signals, at $t_1$ on the galaxy and at $t$ on the Earth, are both of the physical process occurring in a localized small region of space. They are under control of some interaction other than gravity. They are beyond physics of the gravitational field of the Universe. It is thus natural to recognize $\nu_0 = \nu_1$ and the red-shift then means

$$z = \frac{\nu_1 - \nu}{\nu} > 0 \,. \tag{8}$$

If the first wavecrest of the light signal is emitted at time $t_1$ on the galaxy and detected at time $t$ on the Earth, if its next wavecrest is emitted at time $t_1 + \delta t_1$ on the galaxy and detected at time $t + \delta t$ on the Earth, in accordance with Eq.(5), we have

$$-\int_r^0 dr = \int_{t_1}^t \exp(-\sqrt{8\pi\rho G/3}\,t)cdt = \int_{t_1+\delta t_1}^{t+\delta t} \exp(-\sqrt{8\pi\rho G/3}\,t)cdt \,. \tag{9}$$

The second equality in Eq.(9) gives us

$$\int_{t_1}^{t_1+\delta t_1} \exp(-\sqrt{8\pi\rho G/3}\,t)cdt = \int_t^{t+\delta t} \exp(-\sqrt{8\pi\rho G/3}\,t)cdt \,.$$

Integrating this, we find

$$\exp(-\sqrt{8\pi\rho G / 3 t_1})\delta t_1 = \exp(-\sqrt{8\pi\rho G / 3 t})\delta t$$

when $\delta t_1$ and $\delta t$ are small enough. In other words,

$$\frac{v_1}{v} = \exp[\sqrt{8\pi\rho G / 3}(t - t_1)]. \tag{10}$$

Furthermore, the first equality in Eq.(9) gives arise to the relationship among $t_1$, $t$ and $r$ for the light signal,

$$\frac{\sqrt{8\pi\rho G / 3}}{c} r \exp(\sqrt{8\pi\rho G / 3 t}) = \exp[\sqrt{8\pi\rho G / 3}(t - t_1)] - 1. \tag{11}$$

Putting Eqs.(10) and (11) into Eq.(8), we obtain

$$z = \exp[\sqrt{8\pi\rho G / 3}(t - t_1)] - 1, \tag{12}$$

and

$$z = \frac{\sqrt{8\pi\rho G / 3}}{c} r \exp(\sqrt{8\pi\rho G / 3 t}), \tag{13}$$

as the formulas for red-shift of light signals coming from distant galaxies. In Eqs.(12) and (13), the red-shift is described in two ways, in terms of emission time $t_1$ and detecting time $t$ and in terms of detecting time $t$ and distance $r$ of the galaxies from us. The emission time and detecting time are both relative to the SC-moment.

Formula Eq.(13) has more plenty of physical meanings from the experimental point of view. We can rewrite Eq.(13) as

$$cz = H(t)r, \tag{14a}$$

$$H(t) = \sqrt{\frac{8\pi\rho G}{3}} \exp(\sqrt{\frac{8\pi\rho G}{3}}t), \tag{14b}$$

where $H(t)$ is called the red-shift coefficient. The red-shift coefficient is a function of the detecting time. It is positive at all times, from $-\infty$ to $\infty$. That indicates no blue-shift for light signals coming from distant galaxies. The red-shift coefficient increases at all times, too, as time is running from the past to the future. The nowadays red-shift coefficient equals the Hubble parameter $H(t) = H$. But at the SC-moment, it was or will be equal to

$$H(0) = \sqrt{8\pi\rho G / 3}. \tag{15}$$

We have to depend on experimental measurements to determine whether it is "was" or "will be".

### 4. Nowadays cosmological year

We do not have the concept of the age of the Universe because time $t$ in the line element Eq.(3c) can be any value between $-\infty$ and $\infty$. Anyway, to introduce the concept of nowadays cosmological year is appropriate. To estimate it, we recall the data of the mass density of gravity-interacting matter in the Universe and the Hubble parameter [19-21],

$$\rho \approx 0.3 \frac{3H^2}{8\pi G} \text{ and } H = 70\text{~}72 \text{ Km/sec/Mpc.}$$

Putting them into Eq.(14b) and using the equality of $H(t) = H$, one can find

$$Ht = \sqrt{\frac{10}{3}} \ell n \sqrt{\frac{10}{3}},$$

which leads to $t = 15.1$ billion years. This $t$ is positive, so the nowadays cosmological year is around 15.1 billion years after the SC-moment. As for the red-ship coefficient at the SC-moment, we put the data of $\rho$ and $H$ into Eq.(15) and find

$$H(0) = \sqrt{\frac{3}{10}} H = 38.3 \text{ ~ } 39.4 \text{ Km/sec/Mpc.}$$

Around 15.1 billion years ago the red-shift coefficient was little more than half of the nowadays red-shift coefficient.

## 5. "Accelerating Expansion of the Universe"

We already see that the red-shift coefficient increases forever as time is running. The rate of this increase can be found from Eq.(14b),

$$Q(t) \equiv \frac{dH(t)}{dt} = \frac{8\pi\rho G}{3} \exp(\sqrt{\frac{8\pi\rho G}{3}} t). \qquad (16)$$

We call it the increasing rate of red-shift coefficient. It is also positive at all $t \in (-\infty, \infty)$.

In the current standard model of cosmology, the Hubble parameter has the meaning of the expansion rate of the universe [13-18]. If the red-shift coefficient $H(t)$ had the same meaning, Eq.(16) would imply the increasing rate of the expansion rate of the Universe, i.e. an acceleration of the expansion of the Universe. That explains the observed "accelerating expansion of the Universe".

Using the estimated nowadays cosmological year and the data of $\rho$ and $H$ above, we find the nowadays increasing rate of red-shift coefficient or the nowadays value of the "acceleration of the expansion of the Universe",

$$Q(t) \approx 2761 \text{ Km}^2/\text{sec}^2/\text{Mpc}^2.$$

It is very small but detectable, as reported [3-8].

## 6. Concluding remarks

We have derived the formula relating the red-shift of light signals coming from distant galaxies to the distance of these galaxies from us and the time of detecting of these light signals. The red-shift coefficient, instead of the Hubble parameter, has been introduced. The red-shift coefficient is time-dependent. It is positive at all times. It increases forever as time is running from the past to the future. Based on the derived formula, we have further estimated the nowadays cosmological year. It is around 15.1 billion years after the SC-moment. We have also estimated the nowadays increasing rate of red-shift coefficient or the nowadays value of the "acceleration of the expansion of the Universe". It is about 2761 $Km^2/sec^2/Mpc^2$.

However, to us, the red-shift of light signals coming from distant galaxies is not due to the Doppler effect. The root cause of this red-shift lies, not on moving sources or moving detectors of light signals, on the time-evolution of the line element of space-time in the gravitational field of the Universe. This time-evolution affects some, if not all, physical processes occurring in large region of space. This is the influences of the gravitational field of the Universe on these physical processes.

Do we really need the concepts such as the expansion of the Universe, the Big-Bang and the dark energy?

## Acknowledgment

The author greatly appreciates the teachings of Prof. Wo-Te Shen. The author thanks to Dr. J. Conway for his supports and helps.

## References

[1]     Jian-Miin Liu, Reconstructed standard model of cosmology in the Earth-related coordinate system, astro-ph/0505xxx

[2]     E. Hubble, Proc. Natl. Acad. Sci., 15, 168 (1929)

[3]     A. G. Riess et al, Astron. J., 116, 1009 (1998) [astro-ph/9805201]

[4]     S. Perlmutter et al, Astrophys. J., 517, 656 (1999) [astro-ph/9812133]

[5]     R. A. Knop et al, astro-ph/0309368

[6]     P. M. Garnavich et al, Astrophys. J., 509, 74 (1998)

[7]     C. B. Netterfield et al, Astrophys. J., 571, 604 (2002) [astro-ph/0104460]

[8]     C. L. Bennett et al, astro-ph/0302207

[9]     A. Einstein, Autobiographical Notes, in A. Einstein: Philosopheo-Scientist, ed. P. A. Schipp, 3rd edition, Tudor Publishing (New York, 1970)

[10]    Jian-Miin Liu, physics/0208047

[11]    Jian-Miin Liu, A Test of Einstein's Theory of Gravitation: Equilibrium Velocity Distribution of Low-Energy Particles in Spherically Symmetric Gravitational Field, in Fronties in Field Theory, ed. O. Kovras, Nova Science Publishers, Inc. (Hauppauge, NY, 2005) [gr-qc/0206047, 0405048]

[12]    Jian-Miin Liu, Chaos, Solitons&Fractals, 12, 399 (2001); 12, 1111 (2001)

[13]    B. S. Ryden, Introduction to Cosmology, Addison-Wesley (Redwood City, 2002)

[14]    S. Dodelson, Modern Cosmology, Academic Press (New York, 2003)

[15]    S. K. Bose, An Introduction to General Relativity, Wiley & Sons (New York, 1980)

[16]    M. Trodden and S. M. Carroll, astro-ph/0401547

[17]    J. Lesgourgues, astro-ph/0409426

[18]    J. Garcia-Bellido, astro-ph/0502139

[19]    W. L. Freedmann, astro-ph/0202006

[20]    D. N. Spergel et al, astro-ph/0302209

[21]    S. Perlmutter, Physics Today, April, 53 (2003)